\documentclass[twocolumn]{aastex62}

\usepackage{upgreek}
\usepackage{hhline}
\usepackage{amsmath}
\usepackage[LGRgreek]{mathastext}

\graphicspath{{./}{figures/}}

\shorttitle{TRAPPIST-1: Radio obs.}
\shortauthors{Hughes et al.}

\begin{document}

\title{Constraining the Radio Emission of TRAPPIST-1}

\correspondingauthor{Anna Hughes}
\email{ahughes@phas.ubc.ca}

\author[0000-0002-3446-0289]{A. G. Hughes}
  \affiliation{Department of Physics and Astronomy,
  University of British Columbia,
  6224 Agricultural Rd.,
  Vancouver, BC V6T 1Z1, Canada}

\author[0000-0002-0574-4418]{A. C. Boley}
  \affiliation{Department of Physics and Astronomy,
  University of British Columbia,
  6224 Agricultural Rd.,
  Vancouver, BC V6T 1Z1, Canada}

\author[0000-0001-5643-8421]{R. A. Osten}
 \affiliation{Space Telescope Science Institute,
 3700 San Martin Drive,
 Baltimore, MD 21218, USA}
\affiliation{Center for Astrophysical Sciences,
Johns Hopkins University,
Baltimore, MD 21218, USA}

\author[0000-0001-8445-0444]{J. A. White}
  \affiliation{Konkoly Observatory, Research Centre for Astronomy and Earth Sciences, Hungarian Academy of Sciences, Konkoly-Thege Mikl\'os \'ut 15-17, 1121 Budapest, Hungary}




\begin{abstract}

TRAPPIST-1 is an ultracool dwarf (UCD) with a system of 7 terrestrial planets, at least three of which orbit in the habitable zone.  The radio emission of such low-mass stars is poorly understood; few UCDs have been detected at radio frequencies at all, and the likelihood of detection is only loosely correlated with stellar properties. Relative to other low-mass stars, UCDs with slow rotation such as TRAPPIST-1 tend to be radio dim, whereas rapidly rotating UCDs tend to have strong radio emission - although this is not always the case.  We present radio observations of TRAPPIST-1 using ALMA at 97.5 GHz and the VLA at 44 GHz.  TRAPPIST-1 was not detected at either frequency and we place $3 \sigma$ upper flux limits of 10.6 and 16.2 $\upmu$Jy, respectively.  We use our results to constrain the magnetic properties and possible outgoing high energy particle radiation from the star.  The presence of radio emission from UCDs is indicative of a stellar environment that could pose a threat to life on surrounding planets.  Gyrosynchrotron emission, discernible at frequencies between 20 and 100 GHz, is one of the only processes that can be used to infer the presence of high energy particles released during magnetic reconnection events.  Since M dwarfs are frequent hosts of terrestrial planets, characterizing their stellar emission is a crucial part of assessing habitability. Exposure to outgoing high energy particle radiation - traceable by radio flux - can erode planetary atmospheres.  While our results do not imply that the TRAPPIST-1 planets are suitable for life, we find no evidence that they are overtly unsuitable due to proton fluxes. 

\end{abstract}
\section{Introduction}
\indent
M dwarfs are notorious for their high levels of magnetic activity as evidenced by their variability and frequent flares.  Magnetic activity has been observed to increase with decreasing mass for early to mid-M dwarfs \citep{haw_reid,joy}, however, this trend appears to reverse for later spectral types \citep{gunther}.  Ultracool dwarfs - stars and brown dwarfs with spectral type later than M6 - are found to flare more than their $M4-M6$ counterparts \citep{davenport, gunther}.  Unlike early and mid-M dwarfs, main sequence ultracool dwarfs (UCDs) are fully convective \citep{kumar} and unable to generate a magnetic dynamo via the same mechanism(s) as more massive stars \citep{gilman}.  This difference in magnetic dynamo generation could lead to the discrepancy between the magnetic behaviour of UCDs and that of mid-M dwarfs. Although magnetic activity is less common for late-M dwarfs, strong and frequent flares have been observed \citep[e.g.,][]{paudel2, paudel4, paudel1, paudel3, hilton} from active UCDs.

The development of highly sensitive radio telescopes like the \textit{Atacama Large Millimeter Array} (ALMA) and the \textit{Karl G. Jansky Very Large Array} (VLA) has allowed for a small number of UCDs to be detected with quiescent and/or flaring emission. These UCDs have strong radio emission that is often far in excess of values predicted from observational trends for larger magnetically active stars (discussed further in Section \ref{SEC:UCD_RE}).  As with optical flaring and variability, these few but important detections suggest that even the lowest mass stars are capable of producing significant magnetic activity.

While measuring the radio emission of ultracool dwarfs is in itself a way to characterize the behavior of fully convective stars and brown dwarfs, it is also a way to determine how stellar activity could affect the stability of terrestrial planets that orbit UCDs.  M dwarfs are expected to be frequent hosts of terrestrial planets, many of which fall in the habitable zone \citep{dresschar}. Since most stars in the galaxy are M dwarfs, which have the longest lives on the main sequence, most habitable planets could orbit these stars. This assessment of habitability is based on the planetary surface temperature range capable of supporting liquid water at standard pressure, and does not take into account the effects of explosive stellar activity characteristic of at least higher-mass M dwarfs. 

Magnetic reconnection events, detectable at radio frequencies, can release a population of highly energetic particles into the stellar environment. While these events also emit X-ray and UV radiation, the energetic particles may be the most detrimental to planetary atmospheres. Simulations run by \citet{segura} and \citet{tilley} model the effects of outgoing UV and energetic particle radiation from an M3 dwarf flare on the atmosphere of a surrounding Earth-like planet in the habitable zone (0.16 au). The incident UV radiation was based on spectra taken of star AD Leo \citep{hawley}, which was in turn used to estimate the X-ray flux and corresponding energetic ($>10$ MeV) proton flux. Both \citet{segura} and \citet{tilley} found that the ozone in an Earth-like atmosphere was not significantly depleted by UV flares until the addition of energetic particles, which could deplete the ozone column depth by 94\% over the course of 10 years. 

Gyrosynchrotron radiation released during magnetic reconnection events can be used as a tracer of outgoing energetic particles. Reconnection events can manifest in explosive flares and bursting radio emission, or in smaller but ubiquitous reconnection events producing quiescent radio emission \citep{williams2014}. If a star has non-thermal emission present at $\sim 1-100$ GHz frequencies, then potential emission mechanisms can be determined by the location of the peak, the spectral shape, and the flux, depending on the magnetic field strength and electron energy distribution. Observations at higher radio frequencies ($\sim 30-100 $GHz) have the advantage of measuring gyrosynchrotron radiation, which can probe the stellar activity closer to the photosphere and more precisely constrain the size of radio emitting region and magnetic field strength.  While X-ray and $\gamma$-ray observations can determine accelerated particle populations in Solar magnetic events, observations at high radio frequencies where gyrosynchrotron emission is dominant are one of the only ways to constrain accelerated particles in UCDs.  A flux measurement of gyrosynchrotron radiation from a UCD can, in turn, can be used to estimate the energetic proton flux incident on surrounding planets. A non-detection can also be used to place upper limits and constrain particle fluxes.  The UCD TRAPPIST-1 is an interesting target for this work due to its system of 7 rocky planets, which may be threatened by stellar activity. 

TRAPPIST-1 (or 2MASS J23062928-0502285) is a nearby M8 star located  $12.45 \pm 0.02$ pc away from the Sun \citep{kane}.  It has a luminosity of $5.22\pm0.19\times 10^{-4}~ L_{\odot}$, a mass of $0.089\pm0.006~M_{\odot}$, and a radius of $0.121\pm0.003~R_{\odot}$ \citep{van_groot}; this and the absence of lithium absorption \citet{rein_bas} indicate that TRAPPIST-1 is a main sequence UCD (and not a young brown dwarf). Furthermore, \citet{rein_bas} used FeH absorption lines to determine a surface magnetic field of $600^{+200}_{-400}$ G. Additional properties are listed in Table \ref{TBL:T1_props}.

\begin{table}
\caption{Stellar Parameters of TRAPPIST-1 taken from the literature.}
\small
\begin{tabular}{ |p{1.6cm}||p{2.cm}|p{3.7cm}|  }
 \hline
 Parameter & Value & Reference \\ \hhline{|=|=|=|}
 Mass \newline [$M_{\odot}$]&  $0.089 \pm 0.006$ & \citet{van_groot} \\ \hline
 Radius \newline [$R_{\odot}$] & $0.121 \pm 0.003$ & \citet{van_groot} \\ \hline
 T$_{\textrm{eff}}$ \newline [K] & $2516\pm 41$ & \citet{delrez} \\ \hline
 P$_{\textrm{rot}}$ \newline [d] & $3.295 \pm 0.003$ & \citet{vida} \\ \hline
 \textit{v} \,sin\,\textit{i} \newline [$km~ s^{-1}$] & $< 2.0$ & \citet{rein_2018} \\ \hline
 B \newline [G] & $600_{-400}^{+200}$ & \citet{rein_bas} \\ \hline
 L$_\textrm{x}$ \newline [$\frac{erg}{s}$] & $3.8-7.9 \times10^{26} $ & \citet{wheatley} \\ \hline
 Age \newline [Gyr] & $7.6 \pm 2.2$ & \citet{burg_mam} \\ \hline
 d \newline [pc] & $12.43 \pm 0.02$ & \citet{kane} \\ 
 \hline
\end{tabular}
\label{TBL:T1_props}
\end{table}

In 2017, the star was discovered to have a system of seven terrestrial planets \citep{gillon}.  With an effective temperature of $2511\pm37$ K \citep{delrez}, the habitable zone of TRAPPIST-1 is significantly closer to the star than for solar type stars. All orbits of the TRAPPIST-1 planets are within 0.06 au, with a habitable zone encompassing TRAPPIST-1e, f, and g and the closest nominally habitable planet located just 0.02 au from the star. Due to this proximity, the planets are likely tidally locked \citep{gillon}. As such, their slow rotation may reduce the dynamo efficiency, which could in turn inhibit the planets from generating strong magnetic fields. \citet{Greis2004, Greis2007} find this to be the case for hot Jupiters, with the expectation that this will also affect terrestrial planets in the habitable zone of low-mass stars \citep{Khoda}.  Even worse, \citet{garraffo} find that all but the outermost TRAPPIST-1 planets cross into the Alfv\'en surface of TRAPPIST-1's stellar magnetosphere, where the planets are subjected to severe space weather events. Between all of these factors, the TRAPPIST-1 planets are particularly vulnerable to damaging stellar particle radiation.

In this paper we present ALMA and VLA observations of TRAPPIST-1 at 97.5 and 44 GHz respectively, frequencies at which gyrosynchrotron radiation could be present \citep{dulk}, but not likely other types of significant radio emission. Observations of UCDs at such high radio frequencies are scarce. TVLM 513-46546 is the only UCD to be detected in the 45-100 GHz range, with quiescent emission attributed to gyrosynchrotron radiation \citep{Williams_2015}.  The aim of these observations was to determine whether TRAPPIST-1 has comparable radio emission. Scaling the TVLM 513-46546 flux measurements to the size and distance of TRAPPIST-1, the expected emission would be 45 $\mu$Jy at 100 GHz and 60$\mu$Jy at 45 GHz if TRAPPIST-1 is emitting gyrosynchrotron radiation of identical strength. The on-source observation times were chosen in order to achieve a signal-to-noise ratio of 10 for each observation.   The source was not detected at either frequency, which we use to place upper limits on the quiescent flux of TRAPPIST-1 in both cases, constrain the properties of any radio emitting region, and put TRAPPIST-1 in the context of UCD radio emission. The upper level flux was further used to place limits on inferred outgoing energetic proton populations from the star.

\section{Observations}

\begin{figure*}
\begin{center}
\includegraphics[width=15cm]{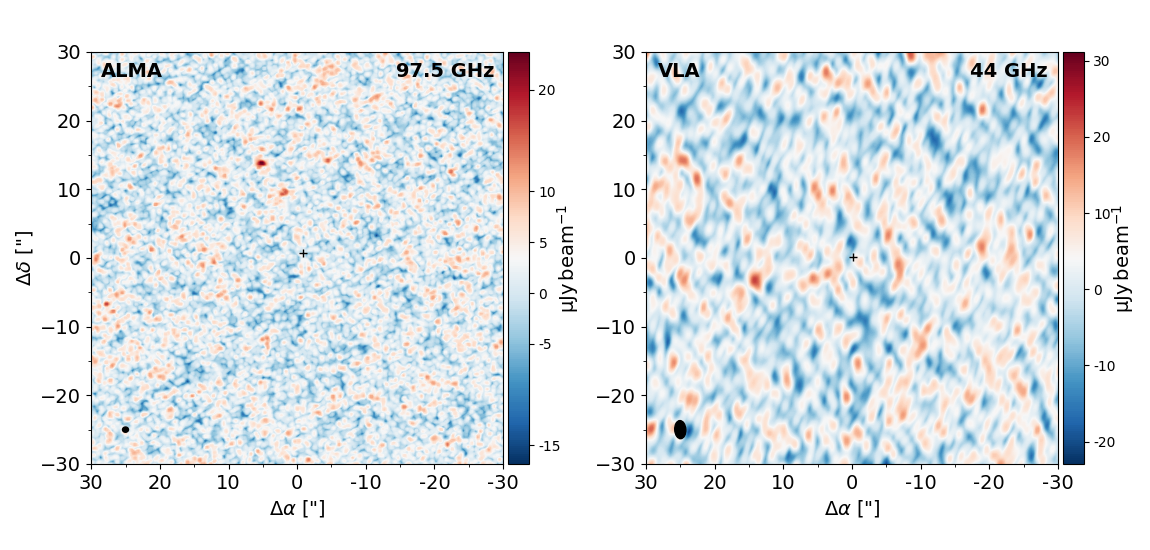}
\caption{Continuum images showing the $60 \times 60$ arcsec region around the positions of TRAPPIST-1 for our 97.5 GHz ALMA (left) and 44 GHz VLA (right) observations.  The crosses at the centers of both images indicate the position of TRAPPIST-1. The synthesized beam is indicated by the black ellipses in the lower left of each image. TRAPPIST-1 was not detected, with RMS sensitivities of $5.34~ \upmu$Jy and $3.52 ~\upmu$Jy in our VLA and ALMA observations, respectively. The two unresolved bright spots in the ALMA image located to the NE of TRAPPIST-1 are likely background object.}
\label{FIG:nondetects}
\end{center}
\end{figure*}

The 97.5 GHz ALMA and 44 GHz VLA observations were centered on TRAPPIST-1 using J2000 coordinates RA = $23^{h} 06^{min} 29.37^{s}$ and $\delta = -05^{\circ} 02' 29.03''$.  The data from both facilities were reduced using the the Common Astronomy Software Applications ({\scriptsize CASA}) pipeline \citep{casa_ref} and are described below.  

\begin{table*}
    \caption{$3 \sigma$ upper flux and radio luminosity limits on TRAPPIST-1. The 6 GHz observations are from \citet{pineda}.}
    \centering
    \begin{tabular}{|l|l|l|l|}
        \hline
        Frequency [GHz] & Flux [$\upmu$Jy] & L$_{\nu,R}$ [erg s$^{-1}$ Hz$^{-1}$] & Ref.\\
        \hhline{|=|=|=|=|}
        $6$ & $< 8.1$ & $<1.5\times10^{12}$ & \citet{pineda}  \\
        $44$ & $<16.2$ &  $<3.0\times10^{12}$ & This work\\
        $97.5$ & $<10.6$ & $<2.0\times10^{12}$ & This work\\
        \hline
    \end{tabular}
    \label{TBL:lims}
\end{table*}

\subsection{ALMA Observations}

The ALMA Cycle 5 observations (ID 017.1.00986.S, PI Hughes) were taken in 8 executions blocks (EBs) from 2018 January 22 to 2018 January 28 for a total of 8.83 hr including overhead and 6.41 hr on-source. There were 43 antennas used with baselines ranging from 15 m to 1397 m. 

Observations were in Band 3 with a total bandwidth of 8 GHz split among 4 spectral windows (SPW). Each SPW has $128\times 15.625$ MHz channels for a total bandwidth of 2 GHz. The SPWs were centred at 90.495 GHz, 92.432 GHz, 102.495 GHz, and 104.495 GHz, giving an effective continuum frequency of 97.50 GHz. The data were reduced using {\scriptsize CASA 4.7.2}, which included WVR calibration; system temperature corrections; flux and bandpass calibration with quasar J0006-0623; and phase calibration with quasar J2301-0158. The precipitable water vapor (PWV) ranged from 1.7 mm to 7.15 mm throughout the observations.

These ALMA 97.5 GHz observations achieve a RMS sensitivity of $3.52~ \upmu \textrm{Jy}~\textrm{beam}^{-1}$ as taken from the CLEANed image. The size of the resulting synthesized beam is $0.835  \times 0.738$ arcsec$^{2}$ at a position angle of $-86.4^{ \circ }$, corresponding to $10$ au at the system distance of 12.45 pc.

\subsection{VLA Observations}

The observations were taken during the VLA Semester 18A (ID VLA-18A-327, PI Hughes) over 4 scheduling blocks (SBs) from 2018 September 4 to 13 for a total of 8.29 hr including overhead and 7.20 hr on-source. Data were acquired with the array in the D antenna configuration, with 26 antennas and baselines ranging from 35 m to 1030 m. 

The instrument configuration used the Q band receiver with a correlator setup consisting of $3968 \times 2.0$ MHz channels for a total bandwidth of 7.936 GHz. Four separate basebands were used with rest frequency centres at 41 GHz, 43 GHz, 45 GHz, and 47 GHz giving an effective continuum frequency of 44.0 GHz. The quasar J2323-0317 was used for gain and phase calibration and quasar 3C48 was used as a bandpass and flux calibrator. Data were reduced using the ({\scriptsize CASA 5.1.2}) pipeline, which included  bandpass, flux, and phase calibrations. 

These VLA 44.0 GHz observations achieve a RMS sensitivity of $5.39 ~ \upmu \textrm{Jy}~\textrm{beam}^{-1}$ as taken from the CLEANed image. The size of the synthesized beam is $2.58 \times 1.53$ arcsec$^{2}$ at a position angle of -86.4$^{ \circ }$. The beam size corresponds to $26$ au at the system's distance.

\subsection{Null detections of TRAPPIST-1}

Our 44 GHz VLA and 97.5 GHz ALMA observations were both non-detections, with $3 \sigma$ upper limits of $16.2 \upmu \textrm{Jy}$ and $10.6 \upmu \textrm{Jy}$ respectively (Table \ref{TBL:lims}).  We confirmed that the phase centre is at the expected location of TRAPPIST-1, taking into account proper motion. 

To ensure that weak variability is not present, we analyzed each observation's infividual scans, which are $\sim 4$ minutes for the VLA observations and $\sim 7$ minutes for the ALMA data. We found no evidence of flaring or variability at median $3 \sigma$ upper limits of 110 $\mu$Jy and 190$\mu$Jy per scan for ALMA and the VLA, respectively.

There are two unresolved bright source candidates in the ALMA image located to the NE of TRAPPIST-1.   They are absent in the VLA observations, and we were not able to identify the object candidates in source catalogues.  The brighter candidate has a flux of $24 \upmu \textrm{Jy}$ and the fainter  has a flux of $16 \upmu \textrm{Jy}$. This corresponds to SNRs of 6.8 and 4.5, respectively.  The significance of this is tested by producing $10^4$ images of Gaussian noise and convolving those images with the synthesized beam.  A $4.5\sigma$ peak was found in 16\% of the realizations, while a $6.8\sigma$ peak did not occur, implying we should expect such a peak less than $0.01 \%$ of the time.  Based on this, the $6.8\sigma$ source is likely real, while the $4.5\sigma$ source may just be noise.

\section{UCD Radio Emission}\label{SEC:UCD_RE}

 
\indent

\begin{figure}
\includegraphics[width=8.35cm]{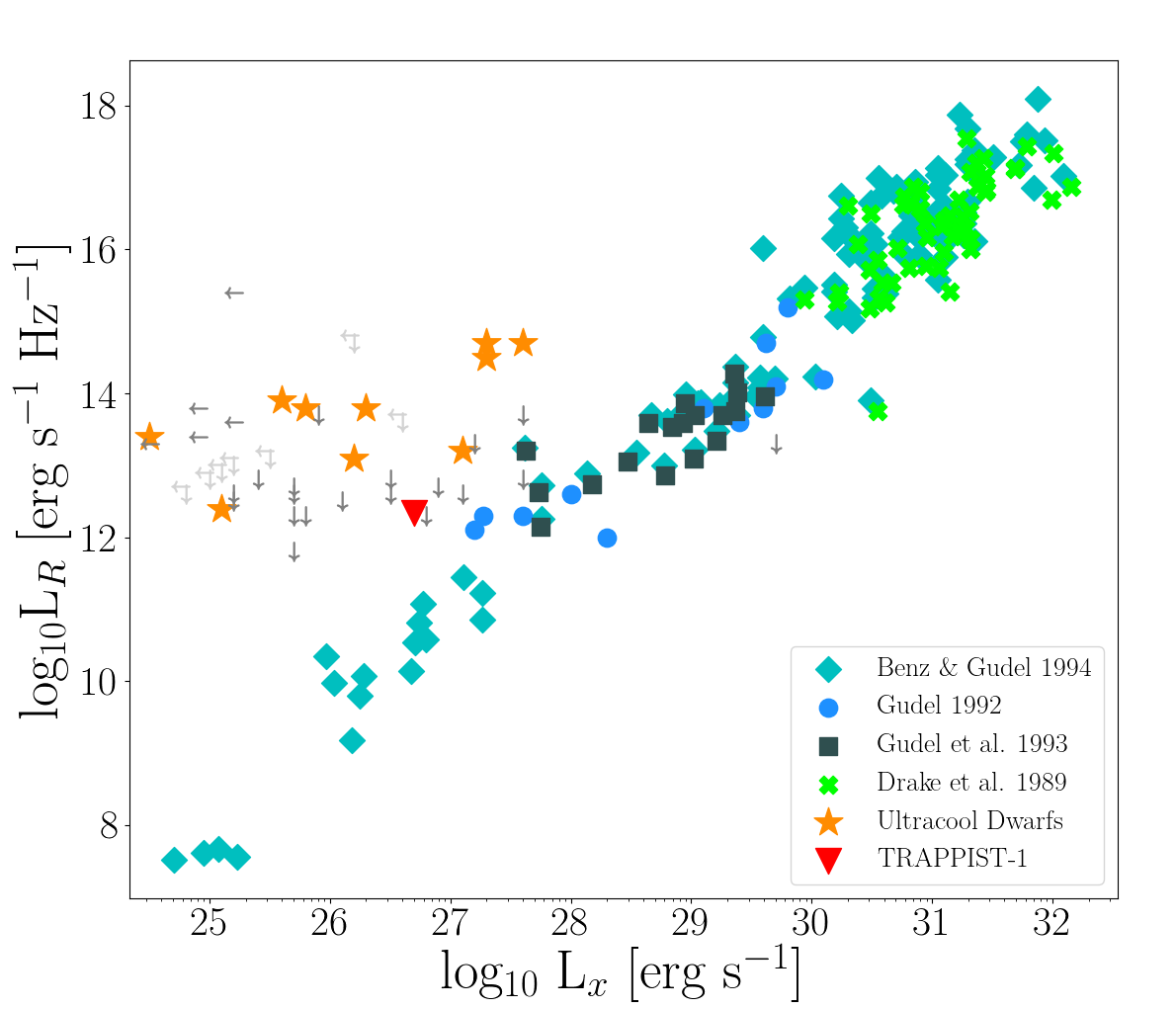}
\caption{The G{\"u}del-Benz relation between X-ray and $\sim 5-9$ GHz radio luminosity. Blue diamond data points represent solar flares from \citet{bg94}, green `x' and blue circles represent magnetically active F, G, \& K stars from \citet{drake} and \citet{gudel92}.  Dark green squares represent mid- to late-M dwarfs in line with the GBR \citep{gudel93}, whereas the orange stars show a population of ultracool dwarfs in violation of the GBR.  Grey arrows show upper-limits on X-ray and/or radio luminosities of UCDs. TRAPPIST-1 is shown along the GBR with a red arrow.  The upper limit in this plot uses the 6 GHz VLA observations by \citet{pineda} and the X-ray luminosity measured by \citet{wheatley}. Uncertainties are not plotted but are typically within $10\%$.}
\label{FIG:gbr}
\end{figure}

While the observations presented in this work are at 44 and 97.5 GHz, we briefly consider lower frequencies to provide context for further discussion of TRAPPIST-1.  The $5-9$ GHz radio luminosity of most magnetically active F through M stars is tightly correlated to the X-ray luminosity, in what is known as the G{\"u}del-Benz relation \citep{gb93,bg94}.  The G{\"u}del-Benz relation (GBR) is well described by a single power law that extends through 10 orders of magnitude (Fig.\,\ref{FIG:gbr}), suggesting a common emission mechanism.  The prevailing model is that radio-emitting non-thermal electrons accelerated in magnetic events heat the coronal plasma, causing the release of soft X-rays \citep{forbrich}. 

This relation is by no means universal; among the small sample of radio emitting UCDs, some have radio luminosity in excess of the GBR-predicted value by up to four orders of magnitude \citep{berger2006}. These deviant radio luminosities, usually determined from approximately 8 GHz fluxes, form a separate branch in the GBR (orange stars in Fig. \ref{FIG:gbr}). It is important to note that while the upper luminosity limits of the null detections in the UCD branch (downward grey arrows) appear the same, this limit is set by telescope sensitivities and these stars may truly be in line with the GBR.  The stars that form the UCD branch are within the mass range expected for full convectivity, which suggest that radio  emission in UCDs may be due to a different magnetic mechanism altogether. Excess radio emission is not present in all UCDs, but is seen more frequently in rapidly rotating UCDs with lower X-ray luminosity.

Indeed, many UCDs follow the GBR. The dark green data points (squares) in Fig. \ref{FIG:gbr} represent a population of M dwarfs including UCDs that fall along the GBR. Recently published 6 GHz observations of TRAPPIST-1 by \citet{pineda} resulted in a null detection. This is shown by the red triangle in Fig. \ref{FIG:gbr}. \citet{pineda} discuss the implications of this null detection extensively, particularly in the context of the electron cyclotron maser instability (ECMI), and we refer the reader to their work for a detailed discussion. In combination with our current work, TRAPPIST-1 has yet to be detected at radio frequencies.


While multiple processes lead to emission at radio wavelengths, only the electron cyclotron maser instability (ECMI) and gyrosynchrotron radiation are thought to be capable of producing the anomalously strong UCD emission seen in the 1-8 GHz regime \citep{pineda, osten2009,ramaty69,burg_put, berger02}, with gyrosynchrotron emission also potentially emitting at the higher frequencies explored here \citep{williams2014}.

ECMI is responsible for the aurorae observed in all solar system giant planets.  Suprathermal electrons follow the electromagnetic currents in the star's magnetic field, and drift in a horseshoe motion around the poles, resulting in radio emission \citep{Wu, Treumann}.  ECMI is characterised by highly circularly polarised radio emission, although ECMI emission could become depolarized during propagation through the UCD magnetosphere.  While ECMI is consistent with observed bursting radio emission of UCDs such as LP 944-20 and DENIS 1048-3956 and the 4-9 GHz emission of some UCDs \citep{berger2006, burg_put, liebert, WilliamsBerger, hallinan}, it cannot account for quiescent flux detected in the $\sim30-100$ GHz regime.

ECMI peaks at the fundamental of the cyclotron frequency, %
\begin{equation}
   \nu_c = 2.8 B_{\textrm{kG}} \,\,\textrm{GHz},
   \label{EQN:peak_freq}
\end{equation}
which for most UCDs is in the GHz regime. Since ECMI emission falls off rapidly for frequencies higher than $\nu_c$, ECMI would require $\sim$ kG magnetic fields strengths to explain any detectable emission.  The measured magnetic field strengths of UCDs is often below this limit \citep{rein_bas}, suggesting ECMI cannot explain the few high-frequency measurements of these objects.  TRAPPIST-1 specifically  has a magnetic field strength of $600^{+200}_{-400}$ G. Unfortunately, with no radio detections so far, we cannot make any definitive statements regarding the potential emission mechanism of TRAPPIST-1.



\section{TRAPPIST-1 in the Context of UCDs}

Despite scarce data, a few trends have been noted in UCD radio emission.  There appear to be two distinct populations of UCDs determined by X-ray luminosity and projected rotation speed: X-ray bright and slowly rotating UCDs tend to be radio dim, while X-ray dim and rapidly rotating UCDs are more likely to have radio emission that exceeds the GBR \citep{williams2014, cook}. All objects that deviate very strongly have $\textit{v} \,sin\,\textit{i} \geq 20~km~s^{-1}$, although the reverse is not true.

This bimodal behaviour is different from early- and mid- M dwarfs, which can exhibit significant activity and correlate with the GBR.  The difference could be related to the change in magnetic field generation at the onset of full convectivity.  Late-type M dwarfs are fully convective, and thus unable to generate magnetic fields via the same mechanisms as solar-type stars, which are thought to rely heavily on the shear between convective and radiative layers \citep{tachocline}.  Prior to the development of highly sensitive radio telescopes, it was not guaranteed that convective stars and brown dwarfs would be capable of producing significant magnetic activity at all \citep{flem, linsk, reid}.  However, in some cases for which the magnetic fields of UCDs have been measured, field strengths can reach up to kG levels \citep{rein_bas}, a thousand times stronger than that of the Sun. While UCDs are unable to generate magnetic fields via the same mechanism as the Sun, there must be some convective dynamo at play.

It is unclear why there are two different populations of UCDs. The UCD branch cannot be explained through variability, as the same break is seen in simultaneous X-ray and radio observations of UCDs \citep{williams2014,berger2008,berger_basri,williams2015,audard,berger2010,berger2009}, including during flares. A few models exist to explain UCD magnetic field generation and corresponding radio emission.  For example, \citet{hallinan2007} and \citet{pineda2017} argue that radio emission due to ECMI requires large, dipolar magnetic fields. In the case of gyrosynchrotron radiation, \citet{williams2014} propose that the magnetic field topology, rather than strength, is responsible for the presence of radio emission.  The divergence in M dwarf behaviour would be the result of two distinct magnetic modes possible in M dwarf populations, where late-type M dwarfs are able to inhabit either mode \citep{morin}. In this bimodal dynamo model, whichever magnetic mode a UCD has is loosely dependent on its rotation rate \citep{mclean}. In the absence of photometric light curves for most UCDs, $\textit{v} \,sin\,\textit{i}$ is used rather than the rotation period. 

Slowly rotating UCDs ($\textit{v} \,sin\,\textit{i} \leq 10~km~s^{-1}$) tend towards axisymmetric dynamos and strong magnetic fields \citep{stelzer}, whereas rapidly rotating UCDs ($\textit{v} \,sin\,\textit{i} \geq 20~km~s^{-1}$) are capable of having either an axisymmetric or non-axisymmetric dynamo and any strength field. Each dynamo creates a distinct field topology, which determines the radio behaviour of the star.  

In their model, \citet{williams2014} posit that UCDs with axisymmetric dynamos have radio emission in agreement with the GBR, whereas the outlying higher radio emission comes from UCDs with weak non-axisymmetric dynamos.  Frequent low-energy magnetic reconnection events due to the tangled multi-polar fields in non-axisymmetric dynamos accelerate electrons along field lines, producing both quiescent and bursting gyrosynchrotron emission at radio frequencies \citep{berger_basri}. 

We find TRAPPIST-1 to be consistent with the trends seen in other UCDs.  With $\textit{v} \,sin\,\textit{i} < 6~km~s^{-1}$ and $L_x \sim 10^{26} \frac{erg}{s}$, it is expected to have negligible radio emission.  The position of TRAPPIST-1 on the GBR is shown in the bottom panel of Fig.\,\ref{FIG:gbr}.  If the emission is below our sensitivity, then TRAPPIST-1 could be in line with the trend for radio quiet UCDs.

\section{Discussion}
\begin{figure}
\begin{center}
\includegraphics[width=0.48\textwidth]{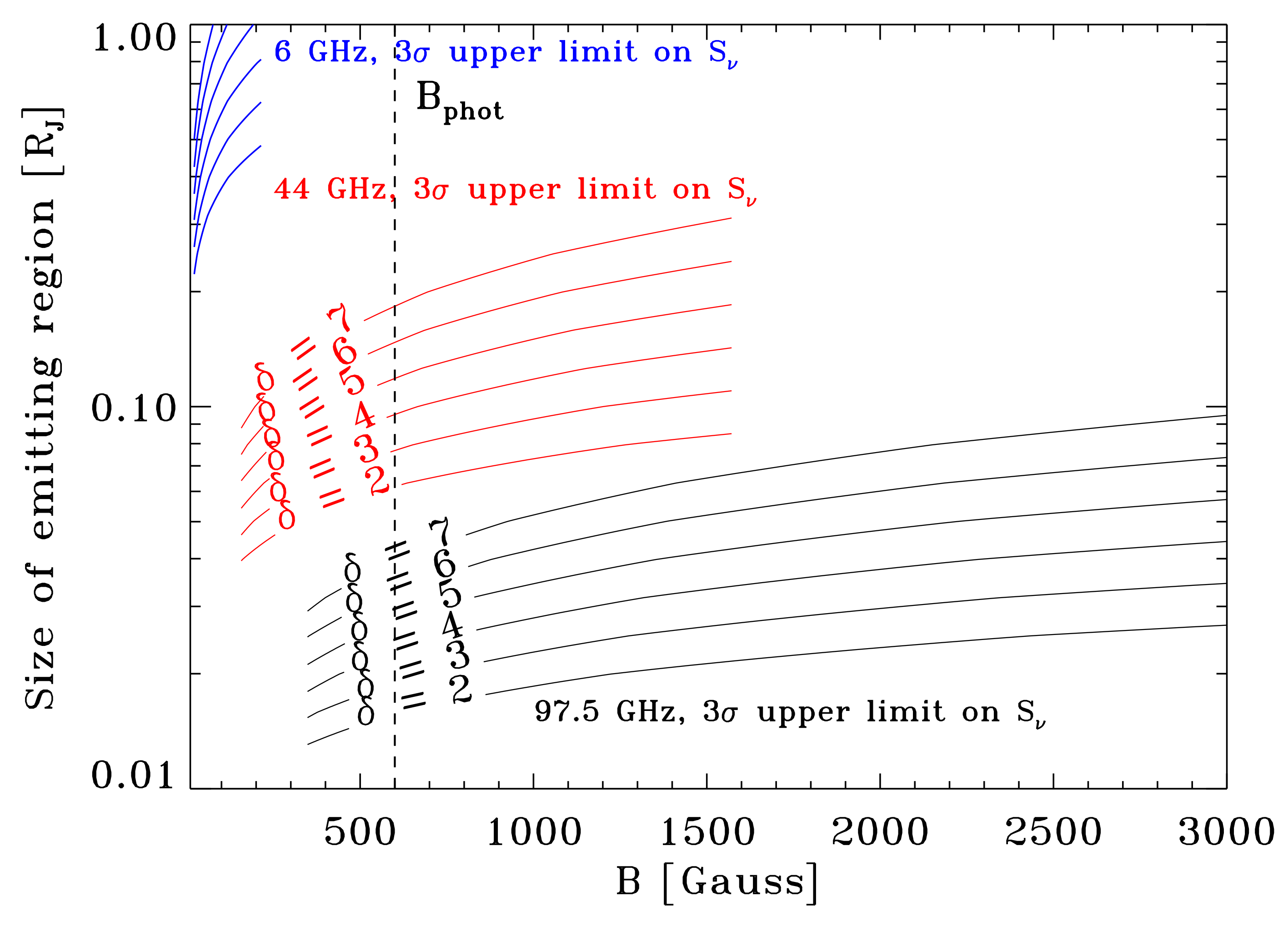}
\end{center}
\caption{Constraints on the size of the emitting region and magnetic field strength for an assumed electron energy index $\delta$ in the optically thin gyrosynchrotron regime.  The upper blue curves are set by the 6 GHz VLA observations presented in \citet{pineda}, while the red and black curves are set by the 44 GHz VLA and 97.5 GHz ALMA observations presented in this work.  The vertical black dotted line shows the magnetic field strength determined by \citet{rein_bas}.}
\label{fig:emit}
\end{figure}

\indent 
While we cannot infer an emission mechanism from only non-detections, we can adopt a particular emission mechanism to calculate some potential constraints on the system. \citet{pineda} focused on ECMI, and we complement that discussion by focusing on gyrosynchrotron emission.
First, following the framework of \citet{white} and 
\citet{osten2009}, we can calculate the size of the emitting region, $x$, in the Rayleigh-Jeans limit:
\begin{equation}
\small
x = 4.5 \times 10^3 \left(\frac{d}{1\rm~pc}\right)\left(\frac{\nu}{1~ \textrm{GHz}}\right)^{-1}\left( \frac{S}{1 ~\upmu \textrm{Jy}}\left(\frac{T_B}{ K}\right)^{-1}\right)^{1/2} R_J
,\label{eqn:regionsize}
\end{equation} 
where $x$ is measured in Jupiter radii $R_J$, d is the distance to the object, $\nu$ is the frequency of observations, $S$ is the flux, and $T_B$ is the brightness temperature of gyrosynchrotron emission.  For the latter, we use the equations in \citet{dulk} appropriate for the optically thin regime. Fig. \ref{fig:emit} shows the range of $x$ values set by each radio observation of TRAPPIST-1 for an assumed electron energy index and magnetic field strength. For a magnetic field of 600 G and electron energy index $\delta = 2$, the emitting region is constrained by our ALMA upper limits to $ \leq 0.02~ R_J$.

\indent
The upper flux limits and size of emitting region can be used to place constraints on outgoing energetic protons during magnetic reconnection events given a few assumptions, which potentially has implications for the habitability of the TRAPPIST-1 planets.  First, the number density of trapped electrons can be constrained for a given upper flux limit and emitting region size. Then, assuming the ratio of trapped electrons to outgoing protons in solar flares holds for UCD reconnection events (although the validity of this is unknown), we can estimate a range of outgoing proton fluxes.

Following the gyrosynchrotron equations arranged by \citet{white} for an assumed electron energy index $\delta=2$, the number density of trapped electrons is given by, \\ %
\begin{flalign}
\scriptsize
N = 4.35 \times 10^3 \Big(\frac{S}{\upmu \textrm{Jy}}\Big) \Big(\frac{B}{G}\Big)^{-1.6} \Big(\frac{R}{R_J}\Big)^{-3} \Big(\frac{d}{pc}\Big)^2 \Big(\frac{\nu}{\textrm{GHz}}\Big)^{0.6} cm^{-3}.
\label{EQN:electron_density}
\end{flalign}

Taking the upper limits set by our ALMA observations, the $3 \sigma$ flux $S = 10.6 ~\upmu \textrm{Jy}$, size of emitting region $R = 0.02 ~R_J$, magnetic field $B=600~ G$, distance $d = 12.45 ~pc$, and frequency $\nu = 97.5$ GHz. The resulting number density of electrons is $N \leq  5 \times 10^8~ cm^{-3}$. Electron energies of gyrosynchrotron radiation range from $\sim$ 10 keV to 100 MeV. For a $\delta = 2$ electron energy index, only 0.001 of the gyrosynchrotron emitting electrons have energies $\geq$ 10 MeV, giving a high-energy electron number density of $N_{10~MeV} \leq 5 \times 10^5 ~cm^{-3}$. If UCD magnetic reconnection events are similar to solar flares, the ratio of electrons to protons is $\approx 10^{3}$ \citep{des_giac}.  
We thus estimate the number density of outgoing $\sim 10 ~MeV$ protons $N_{p^+} \leq 500 ~cm^{-3}$.  Multiplying by the velocity of a $10$ MeV proton ($3 \times 10^{9} ~cm ~s^{-1}$) gives an outgoing proton rate of $\leq 1.5 \times 10^{12} ~cm^{-2} ~s^{-1}$. 

To find a particle flux incident on the TRAPPIST-1 planets, we scale this outgoing proton rate to the distance of the closest habitable planet, TRAPPIST-1e (0.02au).  The distance dependence of the particle flux is not straightforward; Solar System spacecraft demonstrate that the proton flux can have a non-trivial radial profile that is dependent on the specific event.  For example, \citet{lario} find that at distances greater than 1 au the particle flux scales as $r^{-3.3}$ and as $r^{-3}$ at distances less than 1 au. In the absence of similar observations for M dwarfs, however, we scale the outgoing rate by $\frac{R}{r^2}$, where R is the size of the emitting region and r is the semi-major axis. A solid angle term is included to put the result in proton flux units (pfu), where $1 \rm pfu = 1 ~particle ~cm^{-2}~s^{-1}~sr^{-1}$. We use $\pi ~sr$ to represent a flat surface with isotropic incoming radiation.  We find a particle flux incident on the closest habitable planet, TRAPPIST-1e (0.02 au), to be $10^5 pfu$. If TRAPPIST-1 was located along the GBR, the proton flux would be orders of magnitude smaller ($< 500 pfu$).

To put this number into context, solar storms are considered strong by the NOAA (\textit{National Oceanic and Atmospheric Association}) storm radiation scale when the incident protons on Earth reach values of $10^3 ~pfu$, and extend to $\geq 10^5 ~pfu$ for the strongest storms. Simulations ran by \citet{tilley} and \citet{segura} found that an Earth-like planet exposed to UV flares with accompanying incident proton population of $5.9 \times 10^8 ~pfu$ could lose $94 \%$ of its atmospheric ozone over the course of ten years. 

Our results indicate that the TRAPPIST-1 planets are not overtly threatened by catastrophic magnetic processes producing detectable radio emission. This does not, however, guarantee that the TRAPPIST-1 planets are safe from such processes altogether.  While the upper proton flux limit is well below the catastrophic value used by \citet{tilley} and \citet{segura}, it is still within the range considered ``strong" for solar radiation storms on Earth. Smaller scale gyrosynchrotron events such as those seen on the Sun are still possible in the TRAPPIST-1 system.  Bursting rather than quiescent  radio emission may also be possible, but not present during the relatively short on-source timescales of previous observations. The flare rate of TRAPPIST-1 is 0.38 day$^{-1}$ \citep{vida}, meaning that each of our observations only monitored $\sim$10\% of the characteristic timescale of active regions \citep{morris}. Assuming the flares follow a Poisson distribution, then the probability of not detecting a flare in both the ALMA and VLA observations is $\sim$80\%. This difficulty in observing a flare in a single radio observation of TRAPPIST-1 indicates that a long-term radio monitoring campaign is necessary to determine whether TRAPPIST-1 emits gyrosynchrotron radiation during flares.

\section{Summary}

We present 97.5 GHz ALMA and 44 GHz VLA observations of the TRAPPIST-1 system. We find non-detections at both frequencies, and place $3 \sigma$  upper level flux limits of 10.6 $\upmu \textrm{Jy}$ and 16.2 $\upmu \textrm{Jy}$ at 97.5 GHz and 44 GHz, respectively. Analysis of the individual scans showed no signs of variability with median $3 \sigma$ upper limits of 110 $\mu$Jy and 190$\mu$Jy per scan for ALMA and the VLA, respectively. Only 10\% of ultracool dwarfs emit in excess of the  G{\"u}del-Benz relation, with a loose correlation with rotation rate and anti-correlation X-ray luminosity. UCDs with slow rotation rates and high X-ray emission tend to be dim or undetected at radio frequencies, whereas UCDs with high rotation rates and low X-ray emission are more likely to have detectable radio emission. With a slow rotation rate of $3.295 \pm 0.003$ days and high X-ray luminosity of $3.8 - 7.9 \times 10^{26} \frac{erg}{s}$, TRAPPIST-1 conforms to this trend.

Assuming that gyrosynchrotron radiation is the dominant emission mechanism, non-detections can be used to limit the scale of outgoing protons during potential reconnection events for this scenario. Using the upper flux limits set by our observations, along with some basic assumptions about stellar emission mechanisms, we find no evidence that the TRAPPIST-1 planetary system is inherently uninhabitable due to energetic proton fluxes.

\section*{Acknowledgements}

This paper makes use of the following ALMA data: ADS/JAO.ALMA\#2017.1.00986.S. ALMA is a partnership of ESO (representing its member states), NSF (USA) and NINS (Japan), together with NRC (Canada), MOST and ASIAA (Taiwan), and KASI (Republic of Korea), in cooperation with the Republic of Chile. The Joint ALMA Observatory is operated by ESO, AUI/NRAO and NAOJ. The National Radio Astronomy Observatory is a facility of the National Science Foundation operated under cooperative agreement by Associated Universities, Inc.

This work was supported in part by an NSERC Discovery Grant, The University of British Columbia, The Canadian Foundation for Innovation, and the BC Knowledge Development Fund.  This research was undertaken, in part, thanks to funding from the Canada Research Chairs program.  JAW acknowledges support from the European Research Council (ERC) under the European Union's Horizon 2020 research and innovation program under grant agreement No 716155 (SACCRED).

\software{CASA (v5.1.2; McMullin et al. 2007)}

\bibliography{example}



\end{document}